\documentclass[journal=jacsat,manuscript=article]{achemso}
\setkeys{acs}{maxauthors=99}
\setkeys{acs}{doi = true}

\usepackage{graphicx}% Include figure files
\usepackage{dcolumn}% Align table columns on decimal point
\usepackage{bm}% bold math
\usepackage[utf8]{inputenc}
\usepackage[T1]{fontenc}
\usepackage{threeparttable}
\usepackage{chemfig}
\usepackage{mathptmx}
\usepackage{etoolbox}
\usepackage{amsmath}
\usepackage{caption}
\usepackage{subcaption} % for sub figure
\usepackage{url} % to reference external url as hyperlink
\usepackage{sidecap} % figure caption on side instead of at the top or bottom
\usepackage{nameref}
\usepackage{ulem}
\usepackage{amsmath}
\usepackage{lineno}
\usepackage{geometry}
\usepackage{doi}
\usepackage{color}
\usepackage{array}
\usepackage[version=4]{mhchem}
\usepackage[table]{xcolor}
\usepackage{doi}
\usepackage{longtable}
\usepackage{iftex}

\newcommand{\ionp}{CCH$^+$ }
\newcommand{\ion}{CCH$^+$}
\newcommand{\wn}{cm$^{-1}$}
\newcommand{\wns}{cm$^{-1}$ }

\newcommand{\felix}{HFML-FELIX, 6525 ED Nijmegen, The Netherlands}
\newcommand{\imm}{Institute for Molecules and Materials, Radboud University, 6525 AJ Nijmegen, The Netherlands}
\newcommand{\desy}{Photon Science Division, Deutsches-Elektronen-Synchrotron DESY, Hamburg, Germany}
\newcommand{\cologne}{I. Physikalisches Institut, Universit\"{a}t zu K\"{o}ln, 50937 K\"{o}ln, Germany}

\author{Kim Steenbakkers}
\affiliation{\felix}
\alsoaffiliation{\imm}

\author{P. Bryan Changala}
\affiliation{Center for Astrophysics, Harvard and Smithsonian, Cambridge, Massachusetts 02138, United States}
\author{Weslley G. D. P. Silva}
\affiliation{\cologne}

\author{John F. Stanton}
\affiliation{Quantum Theory Project, Departments of Physics and Chemistry, University of Florida, Gainesville, Florida 32611, USA}

\author{Filippo Lipparini}
\affiliation{Dipartimento di Chimica e Chimica Industriale, Università di Pisa, 56124 Pisa, Italy}

\author{J\"{u}rgen Gauss}
\affiliation{Department Chemie, Johannes Gutenberg-Universit\"{a}t Mainz, 55128 Mainz, Germany}

\author{Oskar Asvany}
\affiliation{\cologne}

\author{Gerrit C. Groenenboom}
\affiliation{Theoretical Chemistry, Institute for Molecules and Materials, 6525 AJ Nijmegen, The Netherlands}

\author{Britta Redlich}
\affiliation{\felix}
\alsoaffiliation{\imm}
\alsoaffiliation{\desy}

\author{Stephan Schlemmer}
\affiliation{\cologne}

\author{Sandra Br\"{u}nken}
\affiliation{\felix}
\alsoaffiliation{\imm}
\email{sandra.bruenken@ru.nl}

\title{Experimental proof of strong $\Pi$-$\Sigma$ mixing in the Renner-Teller and Pseudo-Jahn-Teller affected \ionp ($^3\Pi$) ion}

\begin{document}

\begin{abstract}
The ethynyl radical cation, \ionp ($^3\Pi$), offers a unique system for fundamental spectroscopic studies of non-adiabatic effects due to its open-shell linear structure and the presence of a low-lying $^3\Sigma^-$ state, which induces notable perturbations in the (ro-)vibrational spectrum. To probe these effects, we recorded the broadband vibrational spectrum of \ionp from 350–3450~\wns using leak-out spectroscopy. The spectrum reveals a complex splitting pattern in the CCH bending mode, attributed to Renner-Teller and pseudo-Jahn-Teller coupling effects between the $^3\Pi$ and $^3\Sigma^-$ electronic states. A three-state diabatic model, validated here against high-resolution IR data of the CH stretching mode, facilitated assignments within the broadband infrared (IR) spectrum, including an additional $\Pi$ vibronic feature observed in the aforementioned high-resolution spectrum. Our results highlight a pronounced sensitivity of the splitting pattern to the $\Pi$-$\Sigma$ energy gap, with couplings so large that even the zero-point vibrational motion of the bending vibration is sufficient to disrupt the vibronic structure of this ion. This compact ion, with strong coupling effects and high-quality spectroscopic data, serves as an exemplary system for evaluating non-adiabatic models.
\end{abstract}

Non-adiabatic effects - breakdowns of the Born-Oppenheimer approximation that couple electronic and nuclear motion - play a decisive role across a wide range of chemical phenomena. In biological systems they govern their photochemistry \cite{Mai2020} and photostability, for example through conical intersections in nucleic acids and nucleobases \cite{Improta2016}. Moreover, non-adiabaticy has been recognized to play a major role in chemical reactions, with applications ranging from enzyme reactions \cite{Soudackov2014,Zhong2025} to few-atom charge transfer reactions and collisional dynamics important for atmospheric and interstellar environments \cite{Klos2018,Roncero2025,Karman2018}. Despite major theoretical advances, predictive accuracy remains challenging \cite{Takehiro2012}. Direct spectroscopic verifications of non-adiabatic and vibronic coupling effects provide stringent and quantitative tests of theory. Such measurements are especially powerful in few-atom, gas-phase systems as in the present prototypical case, amenable to high-level electronic structure calculations and accessible to experiments free from environmental perturbations and under controlled conditions.
Establishing accurate benchmarks in these well-controlled systems is therefore critical for validating and improving theoretical frameworks that can later be extended to more complex molecular systems.

The ethynyl radical cation (\ion) represents in this context an accurate benchmark system and is of considerable interest from a fundamental spectroscopic perspective, as it is one of the few polyatomic molecules with a $^3\Pi$ electronic ground state for which spectroscopic experimental data are available \cite{Gans2017, Andrews1999,Feinberg2023,deo2004infrared,Steenbakkers_cch}. Treating systems in a $^3\Pi$ state, however, requires the use of sophisticated theoretical approaches to account for the inherent vibronic and angular momentum couplings\cite{Steenbakkers_cch}. In particular, the theoretical treatment of vibrational bending modes is challenging, as symmetry breaking leads to a breakdown of the Born-Oppenheimer approximation. This effect, first described by \citet{renner1934} in 1934 and now known as the Renner-Teller effect, lacked experimental evidence until 1958, when it was first observed in the electronically excited state of the NH$_2$ radical\cite{dressler1959}. This observation provided a benchmark for the effective Hamiltonian model developed by \citet{pople1958theory} in the same year. In 1960, Pople's model was extended to include spin-orbit coupling \cite{pople1960renner}, forming the foundation for the first quantum chemical description of the rovibronic energy levels of a $^3\Pi$ molecule, published in 1962 by \citet{hougen1962vibronic}. Subsequent accumulation of data on molecules in excited $^3\Pi$ states \cite{davies1987infrared,lichten1979fine,simard1988high,ram2010revised,sakamoto2006spectroscopic} led to refinements of Hougen's approach through the development of a standardized effective Hamiltonian\cite{brown1979lambda,brown1977effective,brown1977spin,brown2003rotational,Veseth1971,hirota2012high,he2005renner}. This model now includes spin-orbit coupling, $\Lambda$-doubling, spin-spin coupling, and spin-rotation interaction.  

\hfill

The spectroscopic complexity of \ion, however, extends even beyond these considerations, as demonstrated in the recent theoretical work of \citet{Mehnen2018}. Their calculations at the multireference configuration interaction (MRCI) level of theory using an aug-cc-pV5Z basis set reveal an abundance of low-lying (and crossing) electronic states, potentially giving rise to several additional vibronic coupling effects. Notably, the presence of a $^3\Sigma^-$ state with an adiabatic energy difference of only about 3000 \wns with the ground $^3\Pi$ state may induce (large) coupling effects in addition to the inherent Renner-Teller (RT) coupling, specifically pseudo-Jahn Teller (PJT) coupling, particularly for the vibrational bending mode. The interplay between these effects has been examined in several theoretical studies \cite{liu2010, gorinchoy2011, garcia2012, kayi2012pseudo}, underscoring the necessity of employing three-state non-adiabatic {\it ab initio} models to accurately describe molecules such as the  \ionp ion.

\hfill

From a spectroscopic standpoint, the influence of the low-lying $^3\Sigma^-$ state is anticipated to be highly pronounced in the broadband vibrational spectrum of \ion, as the RT and PJT effects are expected to produce a splitting pattern of the vibrational bending mode, which can span a very large range (up to a few thousand \wn). Additionally, coupling between this low-lying $^3\Sigma^-$ state with the ground $^3\Pi$ state could contribute to a reduction of the expectation value of the orbital angular momentum, thereby affecting the rovibrational line positions and the observed intensities. 

\hfill

Our research groups have now made significant advancements in the experimental investigation of \ion, uncovering the effects of the low-lying $^3\Sigma^-$ state through the acquisition of its broadband vibrational spectrum (350–3450 \wn), high-resolution ro-vibrational spectrum (3065.79–3183.59 \wn), and pure rotational spectrum. Due to the extensive scope of these findings, a comprehensive analysis within a single study was not feasible. As a result, the investigation of \ionp was divided into three parts. \citet{jacob2025} focused on the first detection of \ionp in space and its astrochemical implications, one of our initial motivations to investigate this ion. This detection was based on high-resolution ro-vibrational and pure rotational spectra of \ion, described and analysed with an accurate effective Hamiltonian spectroscopic model in the work by \citet{Steenbakkers_cch} In this paper, we focus on exploring the effects of electronic state mixing as observed in both the broadband and high-resolution ro-vibrational spectra of \ion, interpreting these effects through a three-state diabatic model. This research establishes \ionp as an exceptional benchmark system for evaluating non-adiabatic models.

\begin{figure*}[tb!]
    \centering
    \includegraphics[width=0.9\textwidth]{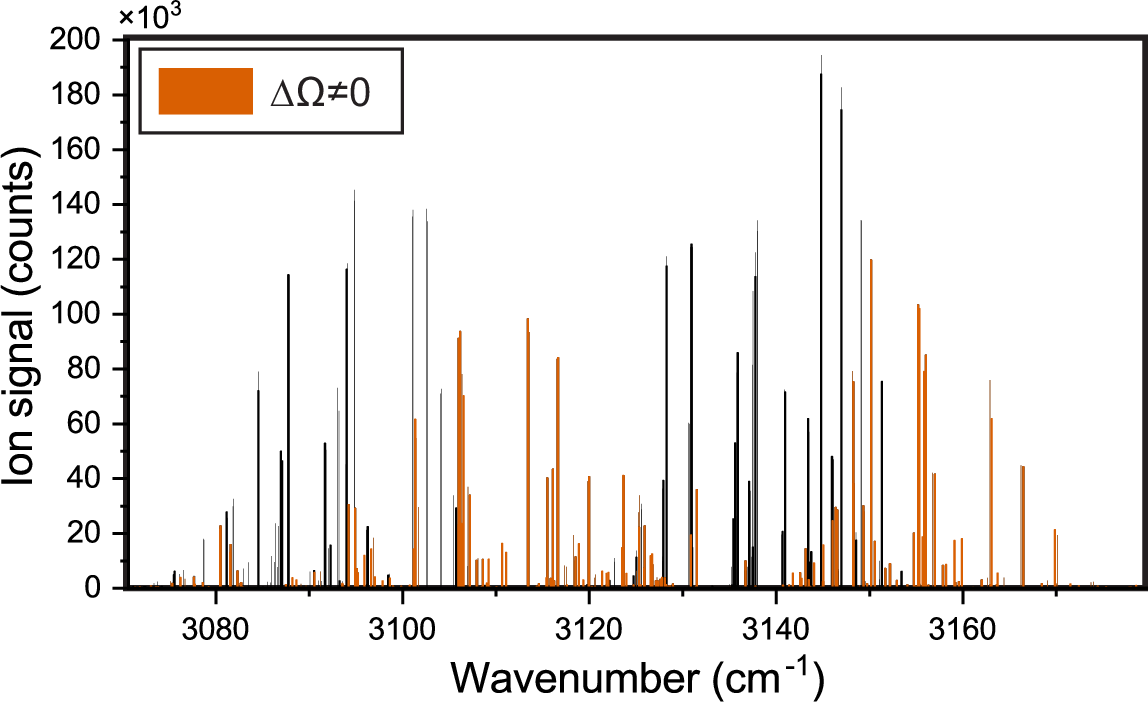}
    \caption{High resolution IR spectrum of \ionp recorded by means of leak-out spectroscopy \cite{Steenbakkers_cch}. The lines highlighted in orange belong to features where $\Delta\Omega \neq$ 0.}
    \label{fig:LOS-letter}
\end{figure*}

\hfill

Recently, we have recorded the high-resolution ro-vibrational spectrum of \ionp in the cryogenic 22-pole ion trap apparatus COLTRAP\cite{Asvany2014} using the leak-out spectroscopy (LOS) method\cite{Schmid2022}, which is in detail described in an accompanying publication \cite{Steenbakkers_cch}. The rotationally resolved spectrum covered the CH stretching fundamental ($\nu_1$) and an additional  vibrational mode of product vibronic $\Pi$-symmetry. The spectrum, shown in Figure \ref{fig:LOS-letter}, shows a multitude of lines consisting of P-, Q-, and R-branches for each of the three different fine-structure components in each of the vibrational bands. As described in detail in \citet{Steenbakkers_cch} the lines were assigned using an iterative method, identifying first transitions in four  strong Q-branches, which could be related to the two distinct vibrational modes based on differences in their corresponding spin-orbit constant, and to the two lowest different fine-structure states based on line intensities. With the help of an effective Hamiltonian model, using the PGOPHER software \cite{Western2019}, corresponding P- and R-branch transition were assigned and added to the effective Hamiltonian analysis, including those of the third fine-structure state. Typically, for $\Pi$-$\Pi$ transitions, only transitions with $\Delta \Omega = 0$ are expected, with $\Omega$ being the projection of the total angular momentum, excluding vibrational angular momentum, on the molecular axis. However, a significant violation of this selection rule was observed, as numerous transitions with $\Delta \Omega = \pm 1$ and $\Delta \Omega = -2$ were detected for both vibrational modes once the $\Delta \Omega = 0$ transitions were assigned (see Figure \ref{fig:LOS-letter}), adding to the complexity of the observed high-resolution spectrum. No transitions for $\Delta \Omega = +2$ were observed due to the relatively low population in the $\Omega=0$ state at the employed nominal trap temperature of 4~K. While the $\Delta \Omega \neq 0$ (cross-$\Omega$) line positions were accurately reproduced by an effective Hamiltonian model described in the work of \citet{Steenbakkers_cch}, the observed intensities relative to the $\Delta \Omega = 0$ transitions were considerably larger than the predicted ones, with the most significant discrepancy occurring in the $\Delta \Omega = -2$ transitions, where the predicted intensities were lower by a factor of approximately 300.

\hfill

The calculated intensities for these cross-$\Omega$ transitions largely stem from the $J_-S_+ + J_+S_-$ term in the effective rotational Hamiltonian, which mediates the transition from Hund's case (a) at low $J$ to case (b) at high $J$. This term induces ($J$-dependent) mixing between the $\Omega$ states, relaxing the $\Delta\Omega = 0$ selection rule, but its effect is relatively small since its magnitude is scaled by the rotational constant, and can therefore not account for the substantial intensities observed in the cross-$\Omega$ transitions. This observation already suggests that the observed transitions are not purely of vibronic $\Pi$-$\Pi$ symmetry in nature but involve significant mixing with the low-lying excited $^3\Sigma^-$ state, which was not considered in the employed effective Hamiltonian \cite{Steenbakkers_cch}. This strong mixing would relax the $\Delta \Omega = 0$ selection rule, thereby explaining the observed deviations, as discussed below.

\hfill

In the present work we have also recorded the broadband spectrum of \ionp in the range of 330–3450 \wns using the LOS method. This approach provides spectra free from line shifts and additional interactions, which are present as a result of the tag used in messenger spectroscopy. Though this method was originally developed for high-resolution rovibrational studies \cite{Schmid2022,Asvany2023,schlemmer2024high,silva2024high}, we have recently demonstrated it to be equally effective for recording the broadband spectra of vibrational bending modes of RT-active molecular ions using a pulsed laser \cite{steenbakkers2024}. The experiments were conducted using the cryogenic 22-pole ion trap instrument, FELion \cite{Jusko2019}, located at the Free Electron Lasers for Infrared eXperiments (FELIX) laboratory \cite{Oepts1995} (see supporting information (SI) for experimental details). 

\hfill

\begin{figure*}[tb!]
    \centering
    \includegraphics[width=0.75\textwidth]{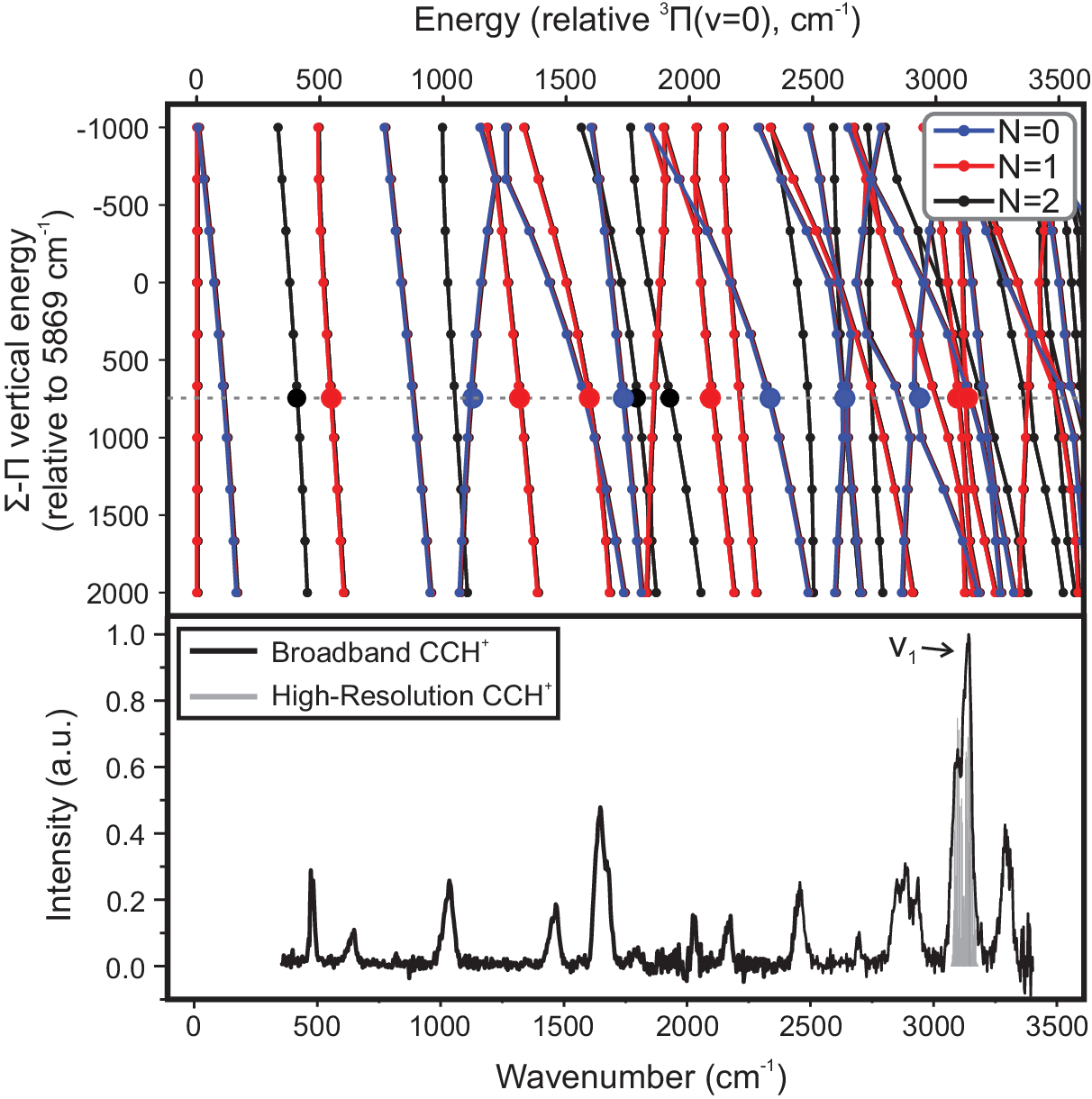}
    \caption{The bottom panel shows the measured broadband spectrum of \ion, recorded by LOS using the FELion instrument, with grey lines representing the high-resolution data (also shown in Figure \ref{fig:LOS-letter}). The top panel displays the calculated rovibronic energies of \ionp as a function of the $^3\Sigma^- - ^3\Pi$ vertical excitation energy for $N = 0-2$ (blue, red, and black, respectively), relative to the vibronic ground-state ($^3 \Pi$, $v = 0$). The $N = 0$ levels are plotted over $N = 1$, which are plotted over $N = 2$, such that vibronic states of $\Sigma$, $\Pi$, and $\Delta$ symmetry effectively also appear as blue, red, and black, respectively. Tentative assignments are highlighted (large dots) at the best estimate shift of $+750$~\wns from the MRCI value, see text and Table \ref{tab:line-pos}.}
    \label{fig:FELIX}
\end{figure*}

The resulting spectrum, presented in the lower panel of Figure \ref{fig:FELIX}, reveals a multitude of vibrational bands. Among these, only the CH stretching mode ($\nu_1$) could be confidently assigned, based on the high-resolution measurements, suggesting that the remaining features predominantly correspond to the RT and PJT splitting components of the CCH bending mode ($\nu_2$), with the exception of the CC stretching feature ($\nu_3$). The assignment of the latter band remains unclear due to the complexity of the splitting pattern and number of vibrational bands observed, and because of its expected large anharmonicity, resulting from the vibronic coupling of the $^3\Pi(A'')$ state with the low lying $^3\Sigma^-$ state. A similar case was observed for the CCH ($^2\Sigma^+$) neutral analogue\cite{stanton2021cc}. 

\hfill

The line positions, intensities, and full-width half maxima are provided in Table \ref{tab:line-pos}, along with a comparison to previous studies and tentative assignments based on calculations described in the next sections. Notably, the feature observed at 3183(1) \wns by \citet{Feinberg2023} using He-droplet spectroscopy, was detected in neither the broadband LOS spectrum nor the high-resolution LOS spectrum. We therefore hypothesize that this feature could correspond to a combination band involving the CH-stretch and a mode associated with the surrounding He atom(s). At this point we also reiterate the necessity of employing a tag-free method, such as LOS, for studying systems affected by vibronic coupling effects. This necessity is exemplified in Figure S1 of the Supporting Information, which shows the fingerprint region of the \ionp spectrum recorded using both LOS and Ne-tagging spectroscopy. It is clear that the Ne-tag completely disrupts the vibronic splitting pattern, which is in line with previous findings for the RT active HCCH$^+$\cite{steenbakkers2024}. Therefore, attempting to interpret tagged spectra as a proxy for the bare ion spectrum appears to be inaccurate for the RT active bending modes. For the RT inactive stretching vibrational modes, however, messenger techniques such as He-droplet isolation and infrared-predissociation can give a good first estimate for the band positions.

\hfill

\begin{table*}[htb!]
\caption{\label{tab:line-pos}The experimental band positions, maximum intensities, and full-width at half-maximum (FWHM) of all observed features in the broadband spectrum of \ionp recorded by means of LOS$^e$, with their (tentative) assignments based on calculated rovibronic energies at a $\Pi$-$\Sigma$ gap of $+750$~\wns offset from the MRCI value.}
 \resizebox{0.9\textwidth}{!}{\begin{threeparttable}
    \begin{tabular}{c c c c c c c c c c}
    \hline
    Band position & Symmetry & Calc. & Dom. Char.** & Max. Int.& FWHM & Ref$^a$ & Ref$^b$ & Ref$^c$ & Ref$^d$\\
    (\wn) &  & (\wn) & $\Lambda(v_1v_2v_3)$  & & (\wn) & (\wn) & (\wn) & (\wn) & (\wn)\\
    \hline\hline
    480(1) & $\Delta$* & 409* & $\Pi(010)\sim59\%$ & 0.26 &	21(1)	&-- &-- &--&--\\
    641(3) & $\Pi$* & 544*	& $\Pi(020)\sim46\%$ & 0.09 &	38(5)	&-- &-- &-- &--\\
    1033(1)	& $\Sigma^+$* & 1109* & $\Pi(010)\sim98\%$ &  0.24 &	46(2)	&-- &-- &-- &--\\
    1462(2)	& $\Pi$* & 1310* & $\Pi(040)\sim29\%$& 0.16 &	42(3)	& --&-- &-- &--\\
    1645(2)	& $\Sigma^-$* & 1574* & $\Pi(011)\sim48\%$ & 0.48 &	42(2)   &-- & 1620(40) &-- & 1832.2(5)\\
    1681(2)	& $\Pi$* & 1592* & $\Pi(001)\sim44\%$ & 0.22 &	27(3)   & --&-- & --&--\\
    1792(5)	&   $\Delta$*   &   1790*   & $\Pi(050)\sim22\%$ & 0.04 &	34(12)	& --&-- & --&--\\
    2029(2)	& $\Delta$* & 1923* & $\Pi(011)\sim27\%$ &  0.13 &	21(3)	& --&-- & --&--\\
    2167(2)	& $\Pi$* & 2091* & $\Pi(001)\sim39\%$ & 0.13 &	31(3)	& --&-- & --&--\\
    2456(2)	& $\Sigma^-$* &  2321* & $\Pi(011)\sim18\%$ & 0.23 &	41(2)   & --&-- & --&--\\
    2695(2)	& $\Sigma^+$* &  2632* & $\Pi(030)\sim91\%$& 0.09 &	19(4) 	& --&-- & --&--\\
    2852(3)	& -- & -- & -- & 0.24 &	46(5)	& --&-- & --&--\\
    2890(2)	& -- & -- & -- & 0.25  &	29(4)	& --&-- & --&--\\
    2932(2)	& $\Sigma^+$* & 2904* & $\Pi(011)\sim 92\%$ & 0.22 &	37(4)	& --&-- & --&--\\
    3091(2)	& $\Pi$    & 3001* & $\Pi(021)\sim 22\%$ & 0.61 &	44(2)	& 3087.5954(2)& -- & 3111(1) &--\\
    3138(2)	& $\Pi$    & 3119 &  $\Pi(100) \sim 65\%$ & 1.00 &	40(1)	& 3136.1220(2)  & --& 3145(1)  &--\\
    --& -- & --&-- &-- & -- & -- & -- & 3183(1) & -- \\
    3294(2)	& -- & -- & -- & 0.39 &	49(1)	& --&-- & --&--\\
    
    \hline 
	
    \end{tabular}
    \begin{tablenotes}
    \item[a] Leak-out spectroscopy from Ref.\ \cite{Steenbakkers_cch}
    \item[b] Slow photoelectron spectroscopy from Ref.\ \cite{Gans2017}
    \item[c] He-droplet spectroscopy from Ref.\ \cite{Feinberg2023}
    \item[d] Ar-matrix spectroscopy from Ref.\ \cite{Andrews1999}
    \item[e] The parameters were determined through fitting to a multi-Gaussian profile. The reported values include the total uncertainties indicated in parentheses.
    \item[*] Tentative assignments
    \item[**] Dominant character. See Table S1 in the SI for a more complete description.
    \end{tablenotes}
    \end{threeparttable}}
\end{table*}

The assignment of the features observed in the broadband spectrum, as well as the interpretation of the ro-vibrational structures in the high-resolution spectrum requires the use of {\it ab initio} calculations based on a three-state diabatic model. These calculations were conducted using a full-dimensional potential energy surface that encompasses the $^3\Pi(A')$, $^3\Pi(A'')$, and $^3\Sigma^-$ states, computed at the MRCI/ANO1\cite{shamasundar2011new,almlof1987general} level of theory with the quantum chemistry package MOLPRO\cite{werner2012molpro,Werner2020}.
The diabatic potential energy and coupling surfaces were derived by fitting the eigenvalues of a quasi-diabatic electronic Hamiltonian expressed as a power series in the vibrational coordinates to the MRCI energies. The rovibronic energies and wavefunctions (neglecting electron spin) were then computed quasi-variationally with the NITROGEN package~\cite{NITROGEN2.1.2} for angular momentum values of $N = 0,1,2$ to determine the level positions of the $\Sigma$, $\Pi$, and $\Delta$ vibronic states. Details of the {\it ab initio} and rovibronic calculations are included in the Supporting Information. 

\hfill

\begin{figure*}[htb!]
    \centering
    \includegraphics[width = 0.8\textwidth]{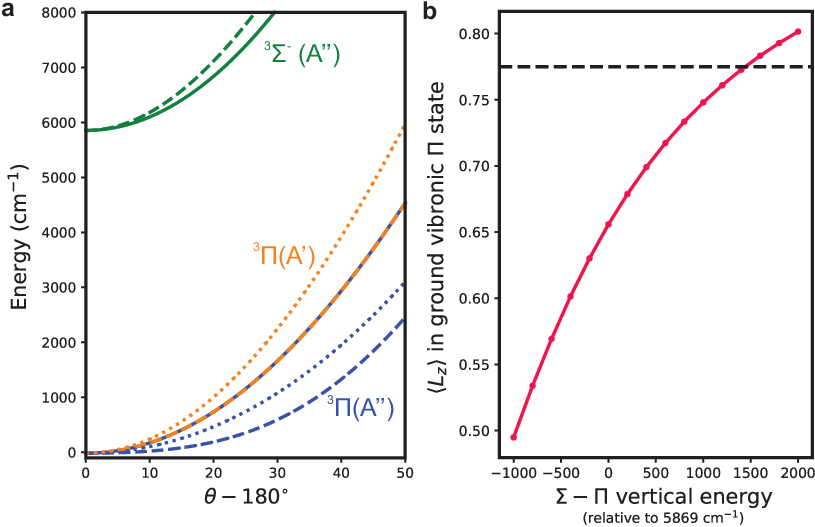}
    \caption{Vibronic interactions in \ion. (a) The potential energy curves for the $^3\Sigma^- (A'')$ (green), $^3\Pi(A')$ (orange), and $^3\Pi(A'')$ (blue) states with respect to the CCH bending angle. The solid curves are the MRCI/ANO1 diabatic surfaces when both RT and PJT interactions are removed, the dotted curves are the two RT-split curves of the $^3\Pi$ states, and the dashed curves are the full adiabatic energies when both RT and PJT interactions are included. 
    (b) Computed expectation value of the projection of the orbital angular momentum  $\langle L_z\rangle$ in the vibronic ground state of \ionp as a function of the $^3\Pi - ^3\Sigma$ vertical excitation energy (red), expressed as the difference from the calculated value 5869 \wns at the MRCI/ANO1 level of theory. The horizontal dashed line is the derived experimental value.}
    \label{fig:quenching}
\end{figure*}

Initially the model was validated with the high-resolution data of the CH stretching mode, the $\Pi$-$\Sigma$ electronic interactions of which we expect to be least affected by RT coupling. A useful approach to examine such orbital angular momentum mixing is to analyze the expectation value of the orbital angular momentum $\langle L_z\rangle$\cite{Jungen1980:RennerTeller}. For a pure $\Pi$ state $\langle L_z\rangle$ is expected to be 1. Any significant deviation from this value indicates mixing between the $\Pi$ and $\Sigma$ states, as $\langle L_z\rangle$ for a pure $\Sigma$ state would be 0. Figure \ref{fig:quenching} illustrates this quenching effect computed for the vibronic ground state within the diabatic model as a function of the $\Pi$-$\Sigma$ vertical excitation energy, here expressed as the difference from the calculated value at the MRCI/ANO1 level of theory of 5869 \wn. It is evident that substantial quenching of the vibronic ground state is expected ($\langle L_z\rangle$=0.5--0.8), even if the $\Pi$-$\Sigma$ gap deviates from the computed value. This observation provides a qualitative explanation for the presence of the cross-$\Omega$ terms. We also note that a small degree of $\langle L_z \rangle$ quenching would be expected even in the absence of $\Pi$-$\Sigma$ vibronic interactions from RT coupling alone, but this contribution should be small. Indeed, we find $\langle L_z \rangle \approx 0.96$ in the vibronic $\Pi$ ground state when only RT coupling is included in the computation, indicating that the quenching in the full model is dominated by $\Pi$-$\Sigma$ interactions.

\hfill

The experimentally derived spin-orbit parameter, $A$, which scales the spin-orbit term ($H_\mathrm{SO} = AL_zS_z$) in the effective Hamiltonian model, provides a direct experimental probe of $\langle L_z\rangle$, because the empirical effective Hamiltonian assumes a vibronic matrix element of $\langle L_z\rangle=1$. Therefore, orbital angular momentum quenching is absorbed into a reduced effective value of the $A$ parameter. The ground state experimental value, $A = -13.831371(70)$ \wn, derived in our accompanying work from the effective Hamiltonian analysis of the ro-vibrational spectrum\cite{Steenbakkers_cch}, is in reasonable agreement with the experimental value reported by \citet{Gans2017}, $|A| = 15$ \wn, obtained by slow photoelectron spectroscopy. However, the calculated value of $A = -18$ \wns by \citet{Mehnen2018} at the equilibrium geometry of the $^3\Pi$ electronic ground state, computed at the MRCI/aug-cc-pVTZ level of theory, overestimates the experimental one. Figure \ref{fig:quenching} can thus be used to estimate the $\Pi$-$\Sigma$ gap by assuming that $\langle L_z\rangle$ is equal to the ratio of the experimentally derived $A$ parameter and the {\it ab initio} value at the equilibrium geometry. In this case, the ratio is $\frac{-13.831371(70)}{-18}=0.78$ (see dashed line in Figure \ref{fig:quenching}), which suggests a $\Pi$-$\Sigma$ gap of $5869+1400=7269$ \wn, indicating that the MRCI/ANO1 calculation underestimates the vertical excitation energy.

\hfill

For comparison, we computed the $\Pi$-$\Sigma$ gap at a higher level of theory, specifically using the partially-spin-restricted coupled cluster approach with single and double excitations augmented by a perturbative treatment of triple excitations\cite{watts1993coupled,Knowles1993a, Knowles1993b} [RCCSD(T)] together with the cc-pCV6Z basis set\cite{wilson1996gaussian,woon1995gaussian}. Higher-order corrections were derived from additional coupled cluster singles, doubles, triples [CCSDT]\cite{noga1987full,watts1990coupled} calculations perturbatively corrected for quadruple excitations [CCSDT(Q)]\cite{bomble2005coupled,kallay2005approximate} and CCSD(T) energies using the cc-pCVTZ basis set\cite{woon1995gaussian,dunning1989gaussian} performed with the CFOUR program package\cite{matthews2020coupled}. This approach should yield very accurate energies at linear geometries, but is not suitable to map a potential energy surface at bent geometries owing to the strong $\Pi-\Sigma$ mixing. This calculation provided a vertical $\Pi$-$\Sigma$ gap of 6302 \wns ($+433$ \wns relative to the MRCI vertical gap) and an adiabatic $\Pi$-$\Sigma$ gap of 3578 \wns ($+822$ \wns relative to the MRCI adiabatic gap), which are in better agreement with the experimental value derived from Figure \ref{fig:quenching} (7269 \wn) than the gap obtained from the MRCI calculations (5869 \wn). Further reported theoretical values (e.g., in Table \ref{tab:line-pos}) are calculated at an energy shift of $+750$~\wns from the MRCI value, chosen as a compromise between the vertical ($+433$~\wn) and adiabatic ($+822$\wn) corrections.

\hfill

\begin{figure*}
    \centering
    \includegraphics[width=\textwidth]{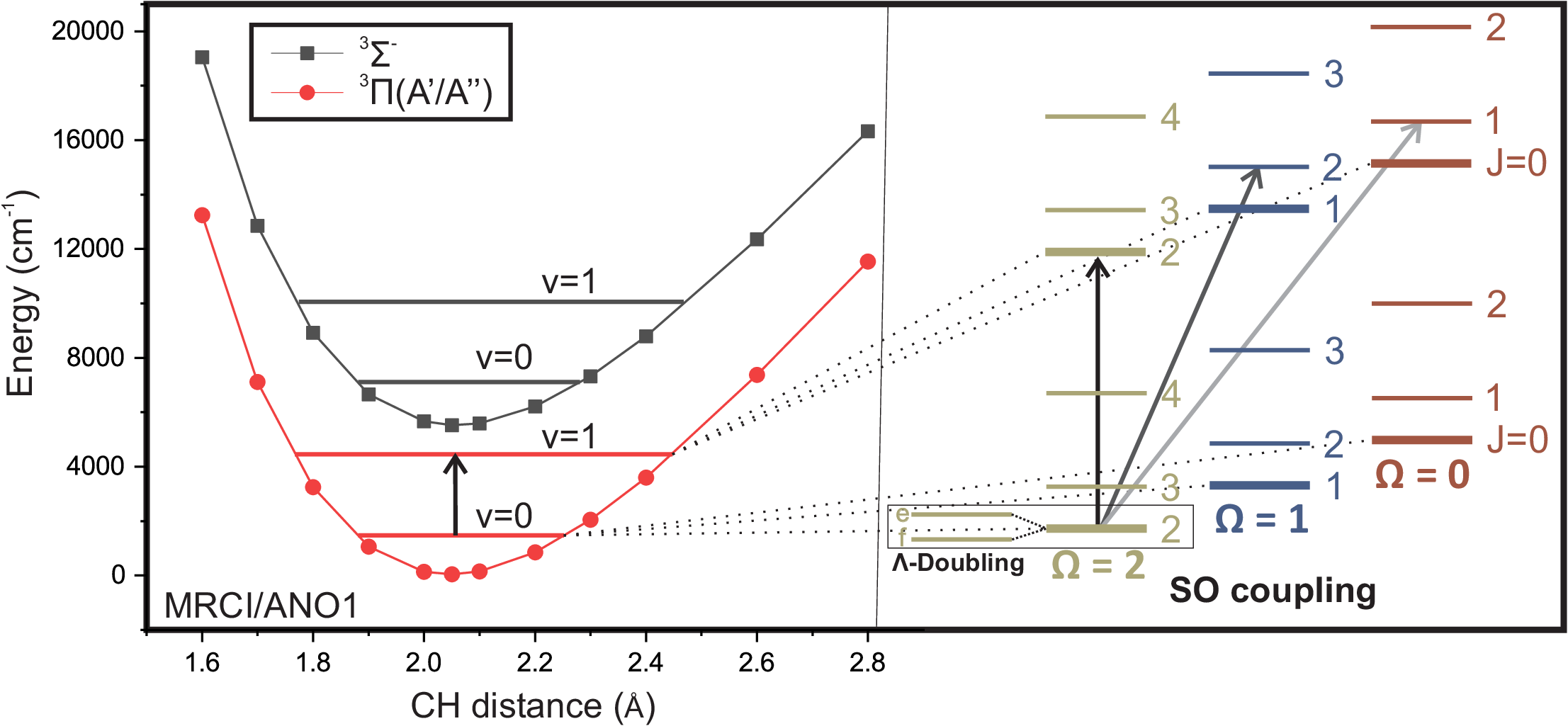}
    \caption{The left panel shows a cut of the potential energy surface along the C-H stretching coordinate of \ion, computed at MRCI/ANO1 level of theory. The energy level splitting as a result of spin-orbit and $\Lambda$-doubling effects is visualized on the right panel (not to scale).}
    \label{fig:term-scheme}
\end{figure*}

The SO constant also shows significant quenching upon excitation to both the $\nu_1$ ($v_1 = 1$) mode and the additional $\Pi$ state, decreasing from $-13.831371(70)$ \wns in the ground state to $-11.87875(38)$ \wns and $-5.95786(31)$ \wns in the respective excited states. This observation suggests a vibrationally dependent quenching of the orbital angular momentum. Examination of the vibrational energy levels on the quasi-parallel potential energy surfaces of the $\nu_1$ mode, as shown in Figure \ref{fig:term-scheme}, indicates that excitation to $v_1 = 1$, calculated at 3119 and observed at 3138(2) and 3136.1220(2)~\wn, in the broadband and high-resolution LOS measurements, respectively, is unlikely to substantially reduce the effective $\Pi$-$\Sigma$ gap. This gap is expected to shift by the difference in the $\nu_1$ fundamental frequencies between the $^3\Pi$ and $^3\Sigma^-$ electronic states, which, within the harmonic approximation, is only approximately 100 \wns and thus insufficient to explain the observed discrepancy. Instead, our computational results reveal a notably lower zeroth-order $v_1 = 1$ component in the corresponding excited-state eigenfunction relative to $v_1 = 0$ in the ground state (see Table S1 in the Supporting Information). Consequently, the calculated ratio of $\langle L_z\rangle$ in the $v = 0$ and $v = 1$ states could serve as an effective diagnostic tool in assigning vibronic features. For the $\nu_1$ mode the calculated ratio of the $\langle L_z\rangle$ values for the $v_1=1$ and $v_1=0$ states is $\frac{0.66}{0.73}=0.90$, which agrees reasonably well with the ratio between the two $A$ constants $\frac{-11.87875(38)}{-13.831371(70)}=0.858827(32)$.

\hfill

Regarding the vibronic feature of $\Pi$ symmetry observed at 3087.59545(18)~\wns in the high-resolution and at 3091 \wns in the broadband LOS spectra, the experimental $\langle L_z \rangle$ value relative to the ground state $\frac{-5.95786(31)}{-13.831371(70)} = 0.430750(52)$ and the respective large shift in the rotational constant of $\Delta B = -0.046516(32)$ \wns provide useful insights for its assignment. Only two additional $\Pi$ vibronic levels are expected to lie in the CH stretching region. These levels are calculated at 3001 \wns and 3091 \wns and have highly mixed vibronic character. For the lower of these two $\Pi$ levels the calculated values, $\frac{0.23}{0.73} = 0.32$ for the change in $\langle L_z \rangle$ and $\Delta B = -0.0577$ \wn, match reasonably well with the experimental values, thus making this the more favorable assignment. This level has an extremely mixed character (see Table S1 in the Supporting Information), and its band origin is highly sensitive to the assumed $\Pi - \Sigma$ energy gap in the diabatic model, as illustrated in the top panel of Figure \ref{fig:FELIX}. Thus, this assignment remains tentative at present.

\hfill

Examination of Figure \ref{fig:FELIX} suggests some tentative assignments for the remaining IR bands observed in the broadband LOS spectrum. For example, the bands observed at 480\,\wns and 641\,\wns appear to belong to the lowest excited $\Delta$ and $\Pi$ vibronic levels, respectively. The $\Sigma^+$ states may provide useful ``anchor points'' because their energies are insensitive to the PJT interactions. The three lowest $\Sigma^+$ vibronic levels, $\Lambda(v_1v_2v_3) = \Pi(010)$, $\Pi(030)$, and $\Pi(011)$, calculated at 1114, 2638, and 2909 \wns are consistent with the bands observed at 1033(1), 2695(2), and 2932(2)\,\wn. 
The state density and congestion at higher vibronic energies make it difficult to securely assign these and other bands. Rotationally resolved spectra of their IR bands would provide definitive vibronic symmetry labels, which, combined with comparisons of spin-orbit quenching and vibrational shifts of rotational constants, would clarify the assignments considerably.

\hfill

Despite these uncertainties in the state-by-state vibronic assignments, some qualitative features of the vibronic structure are clear. The vibronic wavefunctions derived from the 3-state diabatic model demonstrate that excited CC stretching character is highly dispersed amongst multiple vibronic eigenstates. The $v_\mathrm{CC} = 1$ zeroth-order level, for example, is nearly equally fractionated amongst two vibronic eigenstates at ca.\ 1600 \wns and 2100 \wns above the zero-point level, which are split symmetrically about the nominal CC stretch harmonic frequency (1825 \wn). Levels with higher CC stretch character or combined CC stretch-bending character are even more highly mixed (see Table S1 in the Supporting Information) and sensitive to the $\Pi -\Sigma$ electronic gap. The secure identification, via high-resolution spectroscopy, of the two predicted $\Pi$ levels with significant $v_\mathrm{CC} = 1$ character, which we speculatively assign to the bands at 1681 \wns and 2167 \wns in the broadband LOS spectrum, would place useful constraints on the diabatic electronic parameters needed to refine the vibronic assignments.

\hfill

The combined experimental and theoretical spectroscopic study of the ethynyl radical cation (\ion) presented here has yielded significant insights into the intricate non-adiabatic and vibronic coupling effects seen in the corresponding spectra, which arise from the proximity of a low-lying $^3\Sigma^-$ state to the  $^3\Pi$ electronic ground state. The presence of the former results in the violation of the $\Delta\Omega=0$ selection rule, as observed in the high-resolution spectrum of \ion \cite{Steenbakkers_cch}, as shown in Figure \ref{fig:LOS-letter}. Moreover, the broadband vibrational spectrum presented here reveals an intricate splitting pattern caused by a combination of Renner-Teller (RT) and pseudo-Jahn Teller (PJT) coupling effects. Spectral analysis was conducted using computations based on an {\it ab initio} three-state diabatic approach, revealing an extreme sensitivity of the calculated frequencies to the $\Pi$-$\Sigma$ energy gap. Using the measured spin-orbit constants from \citet{Steenbakkers_cch} together with high-level single-point calculations at the coupled cluster level of theory, this gap could be refined, enabling (tentative) assignments of the observed spectral features of the broadband spectrum. While computational predictions provide a tentative framework for understanding these phenomena, further refinements are needed to enhance the accuracy of vibronic state assignments. These refinements could be obtaining high-resolution data on additional vibronic features, some of which are accessible via measurements using quantum cascade lasers\cite{bast2023ro}. Even more promising, LOS has recently been shown to allow recording of rotationally resolved electronic spectra, on the example of another open shell linear ion (HCN$^+$, $^2\Pi$) \cite{Marlton2025,Redondo2026}, opening the way for direct probing of the $^3\Sigma^- \leftarrow X^3\Pi$ electronic transitions and thus the $\Pi$-$\Sigma$ energy gap of \ionp. 
We would like to highlight that for this work the availability of the tag-free LOS method was crucial, as the commonly used messenger (or tagging) spectroscopy disrupts the vibronic structure and complicates the analysis significantly \cite{steenbakkers2024}. LOS proves to be a very general experimental method, not only  for high-resolution ro-vibrational studies, but also for broadband infrared spectroscopy of low-lying bending modes exhibiting the most direct effects of non-adiabatic coupling, and is thus applicable to a wide variety of other potential benchmark systems.

\section*{Acknowledgements}
We gratefully acknowledge the support of Radboud University and of the Nederlandse Organisatie voor Wetenschappelijk Onderzoek (NWO), for providing the required beam time at the FELIX laboratory and the skilful assistance of the FELIX staff. This work is part of the research program “HFML-FELIX: a Dutch Center of Excellence for Science under Extreme Conditions” (with Project No. 184.035.011) of the research program “Nationale Roadmap Grootschalige Wetenschappelijke Infrastructuur,” which is partly financed by the Netherlands Organisation for Scientific Research (NWO). This work has been supported by an ERC Advanced Grant (MissIons: 101020583), and by the Deutsche Forschungsgemeinschaft (DFG) via the Collaborative Research Centre 1601 (project ID: 500700252, sub-projects
C4 and B8) and by DFG SCHL 341/15-1 ("Cologne Center for Terahertz
Spectroscopy"). WGDPS thanks core funding from the University of
Cologne and the Alexander von Humboldt Foundation for funding through a
Postdoctoral Fellowship during the time this work has been carried out. PBC acknowledges support by the US National Science Foundation Award No. AST-2307137.
We would like to thank Ad van der Avoird for inspiring and fruitful discussions about this project.

\section*{Supporting Information}

Sections on the Experimental Methods, the Rovibronic Calculations, a comparison of the Ne-tagging and LOS spectrum (Figure S1), and a table of the calculated vibronic levels (Table S1). 

\bibliography{references}

@article{Redondo2026,
    title = {High resolution overtone spectroscopy of HNC$^+$ and HCN$^+$},
    journal = {Spectrochim. Acta A Mol. Biomol. Spectrosc.}, volume = {349},
    pages = {127359},
    year = {2026},
    issn = {1386-1425},
    doi = {10.1016/j.saa.2025.127359},
    author = {Miguel Jiménez-Redondo and Chiara Schleif and Julianna Palotás
and János Sarka and Hayley Bunn and Petr Dohnal and Paola Caselli and Pavol
Jusko} 
}

@article{karman2018,
  title={O$_2$-O$_2$ and O$_2$-N$_2$ collision-induced absorption mechanisms unravelled},
  author={Karman, Tijs and Koenis, Mark AJ and Banerjee, Agniva and Parker, David H and Gordon, Iouli E and van der Avoird, Ad and van der Zande, Wim J and Groenenboom, Gerrit C},
  journal={Nat. Chem.},
  volume={10},
  number={5},
  pages={549--554},
  year={2018},
  publisher={Nature Publishing Group UK London},
  doi = {10.1038/s41557-018-0015-x},
  URL = {https://doi.org/10.1038/s41557-018-0015-x}
}

@article{Klos2018,
author = {Kłos, Jacek and Bergeat, Astrid and Vanuzzo, Gianmarco and Morales, S{\'e}bastien
B. and Naulin, Christian and Lique, Fran{\c{c}}ois},
title = {Probing Nonadiabatic Effects in Low-Energy C($^3$P$_j$) + H$_2$ Collisions},
journal = {J. Phys. Chem. Lett.},
volume = {9},
number = {22},
pages = {6496-6501},
year = {2018},
doi = {10.1021/acs.jpclett.8b03025},
URL = {  https://doi.org/10.1021/acs.jpclett.8b03025},
eprint = {
        https://doi.org/10.1021/acs.jpclett.8b03025}
}

@article{Improta2016,
author = {Improta, Roberto and Santoro, Fabrizio and Blancafort, Lluís},
title = {Quantum Mechanical Studies on the Photophysics and the Photochemistry of Nucleic Acids and Nucleobases},
journal = {Chem. Rev.},
volume = {116},
number = {6},
pages = {3540-3593},
year = {2016},
doi = {10.1021/acs.chemrev.5b00444},
    note ={PMID: 26928320},
URL = {   https://doi.org/10.1021/acs.chemrev.5b00444},
eprint = { https://doi.org/10.1021/acs.chemrev.5b00444}
}

@Article{Roncero2025,
author ="del Mazo-Sevillano, Pablo and Aguado, Alfredo and Lique, François and Jara-Toro, Rafael A. and Roncero, Octavio",
title  ="Understanding the destruction of CH$^+$ with atomic hydrogen at low temperatures: a non-adiabatic dynamical study",
journal  ="Phys. Chem. Chem. Phys.",
year  ="2025",
volume  ="27",
issue  ="29",
pages  ="15775-15786",
publisher  ="The Royal Society of Chemistry",
doi  ="10.1039/D5CP01718A",
url  ="http://dx.doi.org/10.1039/D5CP01718A"
}

@article{Mai2020,
author = {Mai, Sebastian and González, Leticia},
title = {Molecular Photochemistry: Recent Developments in Theory},
journal = {Angew. Chem. Int. Ed.},
volume = {59},
number = {39},
pages = {16832-16846},
doi = {https://doi.org/10.1002/anie.201916381},
url = {https://onlinelibrary.wiley.com/doi/abs/10.1002/anie.201916381},
eprint = {https://onlinelibrary.wiley.com/doi/pdf/10.1002/anie.201916381},
year = {2020}
}

@article{Zhong2025,
author = {Zhong, Jiayun and Zhu, Qiwen and Soudackov, Alexander V. and Hammes-Schiffer, Sharon},
title = {Hydrogen Tunneling and Conformational Motions in Nonadiabatic Proton-Coupled Electron Transfer between Interfacial Tyrosines in Ribonucleotide Reductase},
journal = {J. Am. Chem. Soc.},
volume = {147},
number = {5},
pages = {4459-4468},
year = {2025},
doi = {10.1021/jacs.4c15756},
    note ={PMID: 39841588},
URL = {  https://doi.org/10.1021/jacs.4c15756},
eprint = { https://doi.org/10.1021/jacs.4c15756}
}

@article{Soudackov2014,
author = {Soudackov, Alexander V. and Hammes-Schiffer, Sharon},
title = {Probing Nonadiabaticity in the Proton-Coupled Electron Transfer Reaction Catalyzed by Soybean Lipoxygenase},
journal = {J. Phys. Chem. Lett.},
volume = {5},
number = {18},
pages = {3274-3278},
year = {2014},
doi = {10.1021/jz501655v},
    note ={PMID: 25258676},
URL = {    https://doi.org/10.1021/jz501655v},
eprint = {  https://doi.org/10.1021/jz501655v}
}

@article{Takehiro2012,
author = {Yonehara, Takehiro and Hanasaki, Kota and Takatsuka, Kazuo},
title = {Fundamental Approaches to Nonadiabaticity: Toward a Chemical Theory beyond the Born–Oppenheimer Paradigm},
journal = {Chem. Rev.},
volume = {112},
number = {1},
pages = {499-542},
year = {2012},
doi = {10.1021/cr200096s},
    note ={PMID: 22077497},
URL = {         https://doi.org/10.1021/cr200096s},
eprint = {  https://doi.org/10.1021/cr200096s}
}

@article{Jacob2025,
    author = {Jakob, Arshia M. and Menten, Karl M. and Brünken, Sandra and Belloche, Arnaud and Wyrowski, Friedrich and Silva, Weslley G. D. P. and Asvany, Oskar and Khan, Sarwar and Kabanovic, Slawa and Steenbakkers, Kim and Groenenboom, Gerrit C. and  Redlich, Britta and Schlemmer, Stephan},
    title = {First detection of CCH$^+$ in the interstellar medium},
    year = {2026}
}

@article{Steenbakkers_cch,
author = {Steenbakkers, Kim and Silva, Weslley G. D. P. and Asvany, Oskar and Groenenboom, Gerrit C. and Jusko, Pavol and Redlich, Britta and Br\"{u}nken, Sandra and Schlemmer, Stephan},
year = {2026},
journal = {J. Phys. Chem. A},
title = {High-resolution ro-vibrational and rotational spectroscopy of the open-shell, linear {CCH}$^+$ ion ($^3{\Pi}$).},
url = {https://doi.org/10.1021/acs.jpca.6c00521},
doi = {10.1021/acs.jpca.6c00521}
}

@article{Knowles1993a,
    author = {Knowles, Peter J. and Hampel, Claudia and Werner, Hans‐Joachim},
    title = {Coupled cluster theory for high spin, open shell reference wave functions},
    journal = {J. Chem. Phys.},
    volume = {99},
    number = {7},
    pages = {5219-5227},
    year = {1993},
    month = {10},
    issn = {0021-9606},
    doi = {10.1063/1.465990},
    url = {https://doi.org/10.1063/1.465990},
    eprint = {https://pubs.aip.org/aip/jcp/article-pdf/99/7/5219/19109440/5219_1_online.pdf},
}

@article{Knowles1993b,
    author = {Knowles, Peter J. and Hampel, Claudia and Werner, Hans-Joachim},
    title = {Erratum: “Coupled cluster theory for high spin, open shell reference wave functions” [ J. Chem. Phys. 99, 5219 (1993)]},
    journal = {J. Chem. Phys.},
    volume = {112},
    number = {6},
    pages = {3106-3107},
    year = {2000},
    month = {02},
    issn = {0021-9606},
    doi = {10.1063/1.480886},
    url = {https://doi.org/10.1063/1.480886},
    eprint = {https://pubs.aip.org/aip/jcp/article-pdf/112/6/3106/19042546/3106_1_online.pdf},
}

@article{davies1987infrared,
  title={Infrared laser spectroscopy of the fundamental band of {A}$^3{\Pi}$ {CO}},
  author={Davies, P. B. and Martin, P. A.},
  journal={Chem. Phys. Lett.},
  volume={136},
  number={6},
  pages={527--530},
  year={1987},
  publisher={Elsevier}
}

@article{lichten1979fine,
  title={Fine structure of 3s, 3d: $^3{\Sigma}$, $^3{\Pi}$, $^3{\Delta}$ complex of {H}$_2$ by {D}oppler-free, laser spectroscopy},
  author={Lichten, W. and Wik, T. and Miller, Terry A.},
  journal={J. Chem. Phys.},
  volume={71},
  number={6},
  pages={2441--2457},
  year={1979},
  publisher={AIP Publishing}
}

@article{deo2004infrared,
  title={Infrared emission studies of the {A}$^3{\Sigma}$---{X}$^3{\Pi}$ electronic transition of the {S}i{C} radical},
  author={Deo, M. N. and Kawaguchi, K.},
  journal={J. Mol. Spectrosc.},
  volume={228},
  number={1},
  pages={76--82},
  year={2004},
  publisher={Elsevier}
}

@article{hougen1962vibronic,
  title={Vibronic and Rotational Energy Levels of a Linear Triatomic Molecule in a $^3{\Pi}$ Electronic State},
  author={Hougen, Jon T.},
  journal={J. Chem. Phys.},
  volume={36},
  number={7},
  pages={1874--1881},
  year={1962},
  publisher={American Institute of Physics}
}

@article{brown1979lambda,
  title={Lambda-type doubling parameters for molecules in ${\Pi}$ electronic states of triplet and higher multiplicity},
  author={Brown, J. M. and Merer, A. J.},
  journal={J. Mol. Spectrosc.},
  volume={74},
  number={3},
  pages={488--494},
  year={1979},
  publisher={Elsevier}
}

@article{werner2012molpro,
  title={Molpro: a general-purpose quantum chemistry program package},
  author={Werner, Hans-Joachim and Knowles, Peter J and Knizia, Gerald and Manby, Frederick R and Sch{\"u}tz, Martin},
  journal={Wil. Int. Rev. Comp. Mol. Sci.},
  volume={2},
  number={2},
  pages={242--253},
  year={2012},
  publisher={Wiley Online Library}
}

@article{almlof1987general,
  title={General contraction of {G}aussian basis sets. {I}. {A}tomic natural orbitals for first-and second-row atoms},
  author={Alml{\"o}f, Jan and Taylor, Peter R.},
  journal={J. Chem. Phys.},
  volume={86},
  number={7},
  pages={4070--4077},
  year={1987},
  publisher={AIP Publishing}
}

@article{shamasundar2011new,
  title={A new internally contracted multi-reference configuration interaction method},
  author={Shamasundar, K. R. and Knizia, Gerald and Werner, Hans-Joachim},
  journal={J. Chem. Phys.},
  volume={135},
  pages={054101},
  number={5},
  year={2011},
  publisher={AIP Publishing}
}

@article{simard1988high,
  title={High-resolution spectroscopy of refractory molecules at low temperature. The e$^3{\Pi}$--a$^3{\Delta}$ band system ($\beta$-system) of {Z}r{O}},
  author={Simard, B. and Mitchell, S. A. and Hendel, L. M. and Hackett, P. A.},
  journal={Faraday Discuss. Chem. Soc.},
  volume={86},
  pages={163--180},
  year={1988},
  publisher={Royal Society of Chemistry}
}

@article{ram2010revised,
  title={Revised molecular constants and term values for the {X}$^3{\Sigma}^-$ and {A}$^3{\Pi}$ states of {NH}},
  author={Ram, R. S. and Bernath, P. F.},
  journal={J. Mol. Spectrosc.},
  volume={260},
  number={2},
  pages={115--119},
  year={2010},
  publisher={Elsevier}
}

@article{sakamoto2006spectroscopic,
  title={Spectroscopic study on vibrational distribution of {N}$_2$ {C}$^3{\Pi}$ and {B}$^3{\Pi}$ states in microwave nitrogen discharge},
  author={Sakamoto, Takeshi and Matsuura, Haruaki and Akatsuka, Hiroshi},
  journal={Jap. J. App. Phys.},
  volume={45},
  number={10R},
  pages={7905},
  year={2006},
  publisher={IOP Publishing}
}

@article{liu2010,
  title={Pseudo {J}ahn-{T}eller versus {R}enner-{T}eller effects in the instability of linear molecules},
  author={Liu, Yang and Bersuker, Isaac B and Zou, Wenli and Boggs, James E},
  journal={Chem. Phys.},
  volume={376},
  number={1-3},
  pages={30--35},
  year={2010},
  publisher={Elsevier}
}

@article{gorinchoy2011,
  title={Jahn--{T}eller, pseudo {J}ahn--{T}eller, and {R}enner--{T}eller effects in systems with fractional charges},
  author={Gorinchoy, Natalia N. and Balan, Iolanta I. and Bersuker, Isaac B.},
  journal={Comp. Theor. Chem.},
  volume={976},
  number={1-3},
  pages={113--119},
  year={2011},
  publisher={Elsevier}
}

@article{garcia2012,
  title={Pseudo {J}ahn-{T}eller origin of bending distortions in {R}enner-{T}eller molecules and its spectroscopic implications},
  author={Garcia-Fernandez, Pablo and Bersuker, Isaac B.},
  journal={Int. J. Quant. Chem.},
  volume={112},
  number={18},
  pages={3025--3032},
  year={2012},
  publisher={Wiley Online Library}
}

@article{kayi2012pseudo,
  title={Pseudo {J}ahn--{T}eller origin of bending instability of triatomic molecules},
  author={Kayi, Hakan and Bersuker, Isaac B. and Boggs, James E.},
  journal={J. Mol. Struct.},
  volume={1023},
  pages={108--114},
  year={2012},
  publisher={Elsevier}
}

@article{steenbakkers2024,
  title={Leak-out spectroscopy as alternative method to rare-gas tagging for the {R}enner--{T}eller perturbed {HCCH}$^+$ and {DCCD}$^+$ ions},
  author={Steenbakkers, Kim and van Boxtel, Tom and Groenenboom, Gerrit C and Asvany, Oskar and Redlich, Britta and Schlemmer, Stephan and Br{\"u}nken, Sandra},
  journal={Phys. Chem. Chem. Phys.},
  volume={26},
  number={3},
  pages={2692--2703},
  year={2024},
  publisher={Royal Society of Chemistry}
}

@article{schlemmer2024high,
  title={High-resolution spectroscopy of the $\nu_3$ antisymmetric {C}--{H} stretch of {C}$_2${H}$_2^+$ using leak-out action spectroscopy},
  author={Schlemmer, Stephan and Plaar, Eline and Gupta, Divita and Silva, Weslley G. D. P. and Salomon, Thomas and Asvany, Oskar},
  journal={Mol. Phys.},
  volume={122},
  number={1-2},
  pages={e2241567},
  year={2024},
  publisher={Taylor \& Francis}
}

@article{silva2024high,
  title={High resolution rovibrational and rotational spectroscopy of {H}$_2${CCCH}$^+$},
  author={Silva, Weslley G. D. P. and Gupta, Divita and Plaar, Eline and Dom{\'e}nech, Jos{\'e} Luis and Schlemmer, Stephan and Asvany, Oskar},
  journal={Mol. Phys.},
  volume={122},
  number={15-16},
  pages={e2296613},
  year={2024},
  publisher={Taylor \& Francis}
}

@article{stanton2021cc,
  title={Why the {CC} stretch in {HCC} is so anharmonic},
  author={Stanton, John F.},
  journal={J. Phys. Chem. A.},
  volume={125},
  number={35},
  pages={7694--7698},
  year={2021},
  publisher={ACS Publications}
}

@article{pople1960renner,
  title={The {R}enner effect and spin-orbit coupling},
  author={Pople, J. A.},
  journal={Mol. Phys.},
  volume={3},
  number={1},
  pages={16--22},
  year={1960},
  publisher={Taylor \& Francis}
}

@article{renner1934,
  title={Zur Theorie der Wechselwirkung zwischen Elektronen - und Kernbewegung bei dreiatomigen, stabf{\"o}rmigen Molek{\"u}len},
  author={Renner, Rudolf},
  journal={Z. Phys.},
  volume={92},
  pages={172--193},
  year={1934},
  publisher={Springer}
}

@article{brown1977effective,
  title={The effective Hamiltonian for the {R}enner-{T}eller effect},
  author={Brown, J. M.},
  journal={J. Mol. Spectrosc.},
  volume={68},
  number={3},
  pages={412--422},
  year={1977},
  publisher={Elsevier}
}

@article{he2005renner,
  title={Renner-{T}eller vibronic analysis for a tetra-atomic molecule. I. {T}he effective {H}amiltonian and matrix elements},
  author={He, Sheng-Gui and Clouthier, Dennis J.},
  journal={J. Chem. Phys.},
  volume={123},
  number={1},
  pages={014316},
  year={2005},
  publisher={AIP Publishing}
}

@article{brown2003rotational,
  title={The rotational dependence of the {R}enner-{T}eller interaction: {A} new term in the effective Hamiltonian for linear triatomic molecules in {$\Pi$} electronic states},
  author={Brown, John M},
  journal={Mol. Phys.},
  volume={101},
  number={23-24},
  pages={3419--3426},
  year={2003},
  publisher={Taylor \& Francis}
}

@article{brown1977spin,
  title={Spin-orbit and spin-rotation coupling in doublet states of diatomic molecules},
  author={Brown, John M. and Watson, James K. G.},
  journal={J. Mol. Spectrosc.},
  volume={65},
  number={1},
  pages={65--74},
  year={1977},
  publisher={Elsevier}
}

@book{hirota2012high,
  title={High-resolution spectroscopy of transient molecules},
  author={Hirota, Eizi},
  volume={40},
  year={2012},
  publisher={Springer Science \& Business Media}
}

@article{dressler1959,
  title={The electronic absorption spectra of {NH}$_2$ and {ND}$_2$},
  author={Dressler, Kurt and Ramsay, Donald Allan},
  journal={Philosoph. Trans. Roy. Soc. London. Ser. A, Math. Phys. Sci.},
  volume={251},
  number={1002},
  pages={553--602},
  year={1959},
  publisher={The Royal Society London}
}

@article{pople1958theory,
  title={Theory of the {R}enner effect in the {NH}$_2$ radical},
  author={Pople, John A and Longuet-Higgins, H Christopher},
  journal={Mol. Phys.},
  volume={1},
  number={4},
  pages={372--383},
  year={1958},
  publisher={Taylor \& Francis}
}

@article{Asvany2014,
   author = {Oskar Asvany and Sandra Brünken and Lars Kluge and Stephan Schlemmer},
   doi = {10.1007/S00340-013-5684-Y/FIGURES/6},
   issn = {09462171},
   issue = {1-2},
   journal = {Appl. Phys. B: Lasers Opt.},
   month = {1},
   pages = {203-211},
   publisher = {Springer},
   title = {{COLTRAP}: {A} 22-pole ion trapping machine for spectroscopy at 4 {K}},
   volume = {114},
   year = {2014},
}

@article{Veseth1971,
title = {Corrections to the spin-orbit splitting in $^2{\Pi}$ states of diatomic molecules},
journal = {J. Mol. Spectrosc.},
volume = {38},
number = {2},
pages = {228-242},
year = {1971},
issn = {0022-2852},
doi = {https://doi.org/10.1016/0022-2852(71)90108-1},
author = {L. Veseth},
}

@Article{Western2019,
author ="Western, Colin M. and Billinghurst, Brant E.",
title  ="Automatic and semi-automatic assignment and fitting of spectra with {PGOPHER}",
journal  ="Phys. Chem. Chem. Phys.",
year  ="2019",
volume  ="21",
issue  ="26",
pages  ="13986-13999",
publisher  ="The Royal Society of Chemistry",
doi  ="10.1039/C8CP06493H",
}

@article{Asvany2023,
   author = {Oskar Asvany and Sven Thorwirth and Philipp C. Schmid and Thomas Salomon and Stephan Schlemmer},
   doi = {10.1039/D3CP01976D},
   issn = {14639076},
   issue = {29},
   journal = {Phys. Chem. Chem. Phys.},
   month = {7},
   pages = {19740-19749},
   publisher = {Royal Society of Chemistry},
   title = {High-resolution ro-vibrational and rotational spectroscopy of {HC}$_3${O}$^+$},
   volume = {25},
   year = {2023},
}

@article{Andrews1999,
   author = {Lester Andrews and Gary P. Kushto and Mingfei Zhou and Stephen P. Willson and Philip F. Souter},
   doi = {10.1063/1.478329},
   issn = {0021-9606},
   issue = {9},
   journal = {J. Chem. Phys.},
   month = {3},
   pages = {4457-4466},
   publisher = {AIP Publishing},
   title = {Infrared spectrum of {CCH}$^+$ in solid argon and neon},
   volume = {110},
   year = {1999},
}

@article{Mehnen2018,
   author = {B. Mehnen and R. Linguerri and S. Ben Yaghlane and M. Mogren Al Mogren and M. Hochlaf},
   doi = {10.1039/C8FD00091C},
   issn = {13645498},
   issue = {0},
   journal = {Farad. Discuss.},
   month = {12},
   pages = {51-64},
   pmid = {30234210},
   publisher = {Royal Society of Chemistry},
   title = {Disentangling the complex spectrum of the ethynyl cation},
   volume = {212},
   year = {2018},
}

@article{Feinberg2023,
   author = {Alexandra J. Feinberg and Swetha Erukala and Cheol Joo Moon and Amandeep Singh and Myong Yong Choi and Andrey F. Vilesov},
   doi = {10.1016/J.CPLETT.2023.140909},
   issn = {0009-2614},
   journal = {Chem. Phys. Lett.},
   month = {12},
   pages = {140909},
   publisher = {North-Holland},
   title = {Isolation and spectroscopy of {C}$_2${H}$^+$ ions in helium droplets},
   volume = {833},
   year = {2023},
}

@article{Gans2017,
   author = {B. Gans and G. A. Garcia and F. Holzmeier and J. Krüger and A. Röder and A. Lopes and C. Fittschen and J. C. Loison and C. Alcaraz},
   doi = {10.1063/1.4973383/280522},
   issn = {00219606},
   issue = {1},
   journal = {J. Chem. Phys.},
   month = {1},
   pages = {11101},
   pmid = {28063431},
   publisher = {American Institute of Physics Inc.},
   title = {Communication: On the first ionization threshold of the {C}$_2${H} radical},
   volume = {146},
   year = {2017},
}

@article{Oepts1995,
   author = {D. Oepts and A. F.G. van der Meer and P. W. van Amersfoort},
   doi = {10.1016/1350-4495(94)00074-U},
   issn = {1350-4495},
   issue = {1},
   journal = {Infrared Phys. Technol.},
   month = {1},
   pages = {297-308},
   publisher = {Pergamon},
   title = {The {F}ree-{E}lectron-{L}aser user facility {FELIX}},
   volume = {36},
   year = {1995},
}

@article{Werner2020,
   author = {Hans-Joachim Werner and Peter J. Knowles and Frederick R. Manby and Joshua A. Black and Klaus Doll and Andreas Heßelmann and Daniel Kats and Andreas Köhn and Tatiana Korona and David A. Kreplin and Qianli Ma and Thomas F Miller and Alexander Mitrushchenkov and Kirk A Peterson and Iakov Polyak and Thomas F Miller III and Guntram Rauhut and Marat Sibaev},
   doi = {10.1063/5.0005081},
   journal = {J. Chem. Phys.},
   pages = {144107},
   title = {The {M}olpro quantum chemistry package},
   volume = {152},
   url = {https://doi.org/10.1063/5.0005081},
   year = {2020},
}

@article{Jusko2019,
   author = {Pavol Jusko and Sandra Brünken and Oskar Asvany and Sven Thorwirth and Alexander Stoffels and Lex van der Meer and Giel Berden and Britta Redlich and Jos Oomens and Stephan Schlemmer},
   doi = {10.1039/C8FD00225H},
   issn = {1359-6640},
   journal = {Faraday Discuss.},
   pages = {172-202},
   title = {The {FEL}ion cryogenic ion trap beam line at the {FELIX} free-electron laser laboratory: infrared signatures of primary alcohol cations},
   volume = {217},
   year = {2019},
}

@article{Schmid2022,
   author = {Philipp C. Schmid and Oskar Asvany and Thomas Salomon and Sven Thorwirth and Stephan Schlemmer},
   doi = {10.1021/acs.jpca.2c05767},
   journal = {J. Phys. Chem. A},
   pages = {8117},
   title = {Leak-Out Spectroscopy, A Universal Method of Action Spectroscopy in Cold Ion Traps},
   volume = {2022},
   url = {https://doi.org/10.1021/acs.jpca.2c05767},
   year = {2022},
}

@misc{NITROGEN2.1.2,
author = {Changala, P. Bryan},
doi = {10.5281/zenodo.7342277},
title = {{NITROGEN, version 2.1.2, https://github.com/bchangala/nitrogen}},
howpublished = {{(https://doi.org/10.5281/zenodo.7342277)}},
year = {2021},
note = {accessed: 2025-01-01}
}

@article{Jungen1980:RennerTeller,
author = {Jungen, Ch. and Merer, A.J.},
doi = {10.1080/00268978000101291},
journal = {Mol. Phys.},
pages = {1--23},
title = {{Orbital angular momentum in triatomic molecules. {I}. {A} general method for calculating the vibronic energy levels of states that become degenerate in the linear molecule (the {R}enner-{T}eller effect)}},
url = {https://www.tandfonline.com/doi/full/10.1080/00268978000101291},
volume = {40},
year = {1980}
}

@article{bast2023ro,
  title={Ro-vibrational spectra of {CC} stretching modes of {C}$_3${H}$^+$ and {HC}$_3${O}$^+$},
  author={Bast, Marcel and B{\"o}ing, Julian and Salomon, Thomas and Thorwirth, Sven and Asvany, Oskar and Sch{\"a}fer, Mathias and Schlemmer, Stephan},
  journal={J. Mol. Spectrosc.},
  volume={398},
  pages={111840},
  year={2023},
  publisher={Elsevier}
}

@article{wilson1996gaussian,
  title={Gaussian basis sets for use in correlated molecular calculations. {VI}. Sextuple zeta correlation consistent basis sets for boron through neon},
  author={Wilson, Angela K. and Van Mourik, Tanja and Dunning Jr., Thom H.},
  journal={J. Mol. Struct.: THEOCHEM},
  volume={388},
  pages={339--349},
  year={1996},
  publisher={Elsevier}
}

@article{woon1995gaussian,
  title={Gaussian basis sets for use in correlated molecular calculations. {V}. Core-valence basis sets for boron through neon},
  author={Woon, David E. and Dunning Jr., Thom H.},
  journal={J. Chem. Phys.},
  volume={103},
  number={11},
  pages={4572--4585},
  year={1995},
  publisher={American Institute of Physics}
}

@article{noga1987full,
  title={The full {CCSDT} model for molecular electronic structure},
  author={Noga, Jozef and Bartlett, Rodney J.},
  journal={J. Chem. Phys.},
  volume={86},
  number={12},
  pages={7041--7050},
  year={1987},
  publisher={American Institute of Physics}
}

@article{watts1993coupled,
  title={Coupled-cluster methods with noniterative triple excitations for restricted open-shell {H}artree--{F}ock and other general single determinant reference functions. {E}nergies and analytical gradients},
  author={Watts, John D. and Gauss, J{\"u}rgen and Bartlett, Rodney J.},
  journal={J. Chem. Phys.},
  volume={98},
  number={11},
  pages={8718--8733},
  year={1993},
  publisher={American Institute of Physics}
}

@article{watts1990coupled,
  title={The coupled-cluster single, double, and triple excitation model for open-shell single reference functions},
  author={Watts, John D. and Bartlett, Rodney J.},
  journal={J. Chem. Phys.},
  volume={93},
  number={8},
  pages={6104--6105},
  year={1990},
  publisher={AIP Publishing}
}

@article{bomble2005coupled,
  title={Coupled-cluster methods including noniterative corrections for quadruple excitations},
  author={Bomble, Yannick J. and Stanton, John F. and K{\'a}llay, Mih{\'a}ly and Gauss, J{\"u}rgen},
  journal={J. Chem. Phys.},
  volume={123},
  number={5},
  year={2005},
  publisher={AIP Publishing}
}

@article{kallay2005approximate,
  title={Approximate treatment of higher excitations in coupled-cluster theory},
  author={K{\'a}llay, Mih{\'a}ly and Gauss, J{\"u}rgen},
  journal={J. Chem. Phys.},
  volume={123},
  number={21},
  pages={214105},
  year={2005},
  publisher={AIP Publishing}
}

@article{dunning1989gaussian,
  title={Gaussian basis sets for use in correlated molecular calculations. {I}. {T}he atoms boron through neon and hydrogen},
  author={Dunning Jr., Thom H.},
  journal={J. Chem. Phys.},
  volume={90},
  number={2},
  pages={1007--1023},
  year={1989},
  publisher={American Institute of Physics}
}

@article{matthews2020coupled,
  title={Coupled-cluster techniques for computational chemistry: The {CFOUR} program package},
  author={Matthews, Devin A. and Cheng, Lan and Harding, Michael E. and Lipparini, Filippo and Stopkowicz, Stella and Jagau, Thomas-C. and Szalay, P{\'e}ter G and Gauss, J{\"u}rgen and Stanton, John F.},
  journal={J. Chem. Phys.},
  volume={152},
  number={21},
  pages={214108},
  year={2020},
  publisher={AIP Publishing}
}

@Article{Marlton2025,
author ="Marlton, Samuel J. P. and Schmid, Philipp C. and Silva, Weslley G. D. P. and Asvany, Oskar and Schlemmer, Stephan",
title  ="High-resolution spectroscopy of [H{,}C{,}N]$^+$ : III. Infrared A$^2\Sigma^+$  ← X$^2\Pi$ electronic transition of HCN$^+$",
journal  ="Phys. Chem. Chem. Phys.",
year  ="2025",
pages  ="-",
publisher  ="The Royal Society of Chemistry",
doi  ="10.1039/D5CP04255K",
url  ="http://dx.doi.org/10.1039/D5CP04255K"
}

\newpage

\section*{TOC Grahic}

\begin{figure*}
   \centering
   \includegraphics[width=\textwidth]{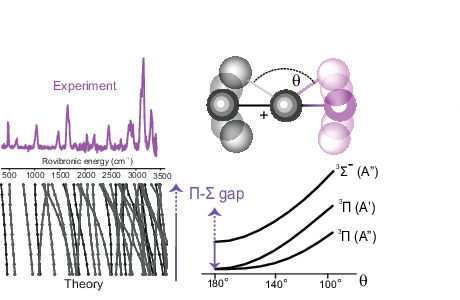}
\end{figure*}

\end{document}